# Dust-free quasars in the early Universe

Linhua Jiang[1], Xiaohui Fan[1,2], W. N. Brandt[3], Chris L. Carilli[4], Eiichi Egami[1], Dean C. Hines[5], Jaron D. Kurk[2,6], Gordon T. Richards[7], Yue Shen[8], Michael A. Strauss[9], Marianne Vestergaard[1,10] & Fabian Walter[2]

[1]Steward Observatory, University of Arizona, 933 North Cherry Avenue, Tucson, Arizona 85721, USA. [2]Max-Planck-Institut für Astronomie, Königstuhl 17, D-69117 Heidelberg, Germany. [3]Department of Astronomy and Astrophysics, Pennsylvania State University, 525 Davey Laboratory, University Park, Pennsylvania 16802, USA. [4]National Radio Astronomy Observatory, PO Box 0, Socorro, New Mexico 87801, USA. [5]Space Science Institute, 4750 Walnut Street, Suite 205, Boulder, Colorado 80301, USA. [6]Max-Planck-Institut für Extraterrestrische Physik, Giessenbachstrasse, D-85748 Garching, Germany. [7]Department of Physics, Drexel University, 3141 Chestnut Street, Philadelphia, Pennsylvania 19104, USA. [8]Harvard-Smithsonian Center for Astrophysics, 60 Garden Street, MS-51, Cambridge, Massachusetts 02138, USA. [9]Department of Astrophysical Sciences, Peyton Hall, Princeton, New Jersey 08544, USA. [10]Dark Cosmology Centre, the Niels Bohr Institute, Juliane Maries Vej 30, DK-2100, Copenhagen O, Denmark.

**The most distant quasars known, at redshifts $z \approx 6$, generally have properties indistinguishable from those of lower-redshift quasars in the rest-frame ultraviolet/optical and X-ray bands[1–3]. This puzzling result suggests that these distant quasars are evolved objects even though the Universe was only seven per cent of its current age at these redshifts. Recently one $z \approx 6$ quasar was shown not to have any detectable emission from hot dust[4], but it was unclear whether that indicated different hot-dust properties at high redshift or if it is simply an outlier. Here we report the discovery of a second quasar without hot-dust emission in a sample of 21 $z \approx 6$ quasars. Such apparently hot-dust-free quasars have no counterparts at low redshift. Moreover, we demonstrate that the hot-dust abundance in the 21 quasars builds up in tandem with the growth of the central black hole, whereas at low redshift it is almost independent of the black hole mass. Thus $z \approx 6$ quasars are indeed at an early evolutionary stage, with rapid mass accretion and dust formation. The two hot-dust-free quasars are likely to be first-generation quasars born in dust-free environments and are too young to have formed a detectable amount of hot dust around them.**

More than 40 quasars have been discovered at redshifts $z \approx 6$ (refs 5, 6); at this epoch, the Universe was less than one billion years old. They harbour black holes with masses higher than $10^8$ solar masses ($10^8\,M_\odot$) and emit radiation at about the Eddington limit[2,7]. According to unification models of active galactic nuclei, the black hole accretion disk is surrounded by a dusty structure[8,9]. A significant fraction of the quasar's ultraviolet and optical radiation is absorbed by the dust and is re-emitted at infrared wavelengths. In particular, the hottest dust is directly heated by the central engine and produces near-infrared (NIR) emission[10–12]. Radiation from hot dust is observed to be a ubiquitous feature among quasars, and is strong evidence for these unification models.

In order to study the properties of hot dust at high redshift, we obtained deep infrared photometry of 21 $z \approx 6$ quasars using the Spitzer Space Telescope (referred to here as Spitzer). The quasar sample spans a redshift range of $5.8 < z < 6.4$ and a luminosity range of $-27.9 < M_{1450} < -25.4$, where $M_{1450}$ is the absolute AB magnitude of the continuum at rest-frame 1,450 Å. A cosmology with $H_0 = 70$ km s$^{-1}$ Mpc$^{-1}$, $\Omega_m = 0.3$ and $\Omega_\Lambda = 0.7$ is adopted. The observations were carried out in the four IRAC channels (3.6, 4.5, 5.8 and 8.0 μm), the IRS Peak-Up Imaging (PUI) blue band (15.6 μm) and the MIPS 24 μm band. These bands correspond to a rest-frame wavelength range of $0.5 \leq \lambda_0 \leq 3.5$ μm and trace the disk and hot dust radiation for $z \approx 6$ quasars. Here we use $\lambda$ and $\lambda_0$ to denote observed and rest-frame wavelengths, respectively. All the quasars except SDSS J000552.34−000655.8 (hereafter J0005−0006; we use JHHMM ± DDMM for brevity) and J0303−0019 were detected with high signal-to-noise ratios in all the Spitzer bands used. However, J0005−0006 was not detected at 15.6 and 24 μm, and J0303−0019 was not detected at 24 μm. Note that J0005−0006 was not detected in our earlier, much shallower Spitzer 24 μm imaging[4].

Figure 1 shows the Spitzer spectral energy distributions (SEDs) of six representative objects in our sample. Their continuum shapes at the wavelengths of the IRAC bands ($0.5 \leq \lambda_0 \leq 1$ μm) are consistent with the average SED of low-redshift type 1 (unobscured) quasars[13]. At 15.6 and 24 μm ($\lambda_0 \approx 2$ and 3.5 μm), however, the quasar SEDs show wide diversity. For instance, J1250+3130 has strong NIR excess, whereas J0005−0006 and J0303−0019 are undetected in the MIPS 24 μm band. In a type 1 quasar, the radiation at $\lambda_0 < 1$ μm is mostly from the accretion disk; at longer wavelengths of a few micrometres, hot-dust radiation dominates over the disk radiation. We use a simple model to fit the Spitzer SEDs, consisting of a power-law disk component and a hot-dust blackbody with a temperature of 1,200 K (ref. 14; Fig. 1). Most of the quasars in our sample show strong NIR emission excess that we interpret as due to hot dust, but J0005−0006 and J0303−0019 do not show any detectable emission from hot dust. These two quasars have the narrowest emission lines (line widths ~2,000 km s$^{-1}$) of the objects in the sample[7,15]. They are not obviously luminous counterparts of narrow-line Seyfert 1 galaxies, because such galaxies tend to have NIR emission similar to or stronger than that of ordinary Seyfert galaxies[16].

NIR spectroscopy[7,15] and our IRAC photometry show that J0005−0006 and J0303−0019 have normal rest-frame ultraviolet/optical SEDs and broad emission lines (although among the narrowest of the sample). It is very unlikely that emission from hot dust is obscured by optically thick cooler dust at larger scales, as the broad-line region is seen along the line of sight. Furthermore, these objects are undetected at 250 GHz (r.m.s. noise of 0.5 mJy), ruling out cool dust emission[17,18]. J0005−0006 was also observed by the Chandra X-ray satellite. Its X-ray emission is consistent with a typical type 1 quasar, indicating no unusual obscuration[3]. Although hot-dust emission was thought to be a ubiquitous feature in quasars, it apparently does not exist in these two $z \approx 6$ quasars.

Indeed, hot-dust-free quasars like J0005−0006 and J0303−0019 do not have counterparts at low redshift. Figure 2 illustrates the hot-dust abundance for our $z \approx 6$ sample, as well as for an unbiased sample of 362 low-redshift quasars with rest-frame NIR flux measurements in the literature[12,13,19–22].

Here we use the ratio of the rest-frame NIR-to-optical (3.5 μm to 5,100 Å) emission as an indicator of the hot-dust abundance. The hot-dust abundances for most objects are close to 1, with little dependence on luminosity over four orders of magnitude or on redshift over a range of $0 < z < 6$. The tight correlation between the two luminosities suggests that hot dust is directly heated by quasar activity. However, J0005−0006 and J0303−0019 do not show any hot-dust emission and thus significantly deviate from the correlation. No known quasars at low redshift have hot-dust abundances as low as these $z \approx 6$ quasars. This is the first concrete evidence that SEDs from the quasar central engine have evolved significantly at high redshift: accretion disks in most $z \approx 6$ quasars may have reached maturity, but dust structures are still evolving rapidly.

Quasars are powered by mass accretion onto central black holes and hot dust is directly heated by quasar activity, so the black hole masses may provide useful clues to the puzzle as to why J0005−0006 and J0303−0019 do not have hot dust. Black hole masses in distant quasars can be estimated using mass scaling relations based on broad emission line widths and continuum luminosities[23]. Figure 3 presents the hot-dust abundance as a function of the estimated black hole mass for quasars at redshifts from $z < 0.7$ to $z \approx 6$. At low redshift, the hot-dust abundance is almost independent of the black hole mass. This explains why we do not find low-redshift quasars like J0005−0006 and J0303−0019, because at low redshift dust has supposedly been recycled between galaxies and the intergalactic medium, and quasar host galaxies are rich in dust even before the central black hole first undergoes rapid accretion. At $z \approx 6$, however, the hot-dust abundance is roughly proportional to the black hole mass, indicating that the two grow at about the same rate. The two hot-dust-free quasars with the lowest hot-dust abundances have the smallest black hole masses ($2$–$3 \times 10^8$ $M_\odot$) and highest Eddington luminosity ratios (~2) in the $z \approx 6$ sample. This suggests that they are in an early stage of quasar evolution with rapid mass accretion, but are too young to have formed a detectable amount of hot dust around them.

In the local Universe, most dust in the interstellar medium is produced by low- and intermediate-mass AGB stars, which develop 0.5–1 billion years after the initial starburst. At $z \approx 6$, the age of the Universe was less than one billion years, and the first star formation probably occurred less than half a billion years earlier. Therefore, other dust production mechanisms may play an important role at high redshift. For example, quasar activity itself can efficiently produce dust through outflowing winds[24]. The strong correlation between the hot-dust abundance and black hole mass in $z \approx 6$ quasars suggests that hot dust in these quasars may be produced in this way. Supernovae are another alternative source of dust at high redshift. For example, the colours of a reddened quasar at $z = 6.2$ and the host galaxy of GRB 050904 at $z = 6.29$ cannot be explained by low-redshift extinction curves, but are well fitted by dust originating from supernovae[25,26]. Although the reddening dust in these two cases would be on larger scales than the hot dust considered in our study, it does support the notion that dust properties can show strong evolution at high redshift.

Until now, multi-band observations have shown that $z \approx 6$ quasars are well-evolved objects in the early Universe as they are indistinguishable from lower-redshift quasars in the wavebands observed. However, our Spitzer observations, especially the discovery of two hot-dust-free quasars, indicate that $z \approx 6$ quasars are indeed young and their hot dust evolves rapidly. The strong relation between the hot-dust abundance and black hole mass shows an evolutionary relation at $z \geq 6$ (Fig. 3): small black holes are growing rapidly, with hot dust building up at similar rates, towards a mature structure with a hot-dust abundance close to one. We thus suggest that J0005−0006 and J0303−0019 are living in extremely young environments; they are likely to be first-generation quasars, which have been born in a dust-free medium and are in their earliest evolution stage. Dust is being produced for the first time in those environments.

**Acknowledgements** This work is based on observations made with the Spitzer Space Telescope, which is operated by the Jet Propulsion Laboratory, California Institute of Technology under a contract with NASA. Support for this work was provided by NASA through an award issued by JPL/Caltech. X.F. acknowledges support by NSF AST, a Packard Fellowship for Science and Engineering, and a John Simon Guggenheim Memorial Fellowship. W.N.B. was supported by the NASA ADP program. C.L.C. thanks the Max-Planck-Gesellschaft and the Humboldt-Stiftung for support through the Max-Planck-Forschungspreis. J.D.K. thanks the DFG for support via German-Israeli Project Cooperation. The Dark Cosmology Centre is funded by the Danish National Research Foundation.

**Author Contributions** L.J. and X.F. designed the project, reduced and analysed the data, and prepared the manuscript; W.N.B., M.A.S. and F.W. performed statistics and edited the manuscript; C.L.C., E.E., D.C.H. and G.T.R. prepared observations; J.D.K. analysed NIR spectra of two hot-dust-free quasars; Y.S. and M.V. measured black hole masses. All authors helped with the scientific interpretations and commented on the manuscript.

**Competing interests statement** The authors declare that they have no competing financial interests.

**Correspondence** Correspondence and requests for materials should be addressed to L.J. (e-mail: ljiang@email.arizona.edu).

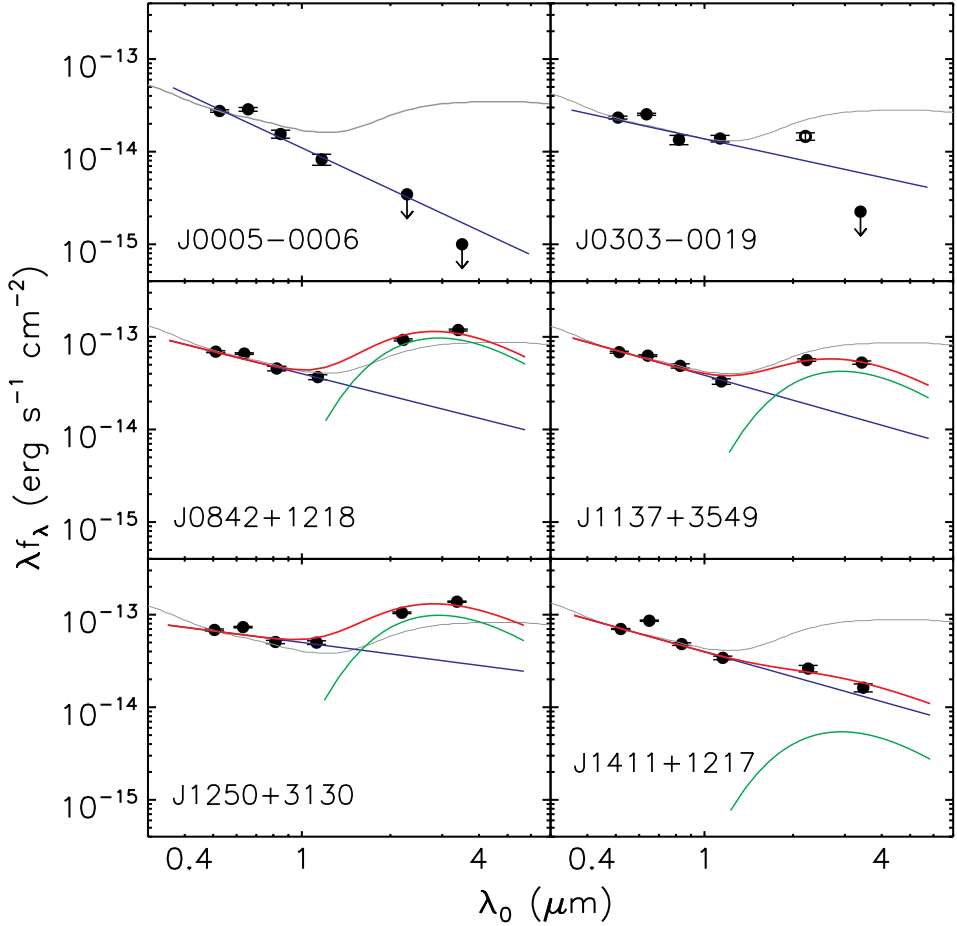

Figure 1. Spitzer SEDs of z ~ 6 quasars. The circles show the rest-frame SEDs of six representative quasars from our sample in the four IRAC channels, the IRS PUI blue band (15.6 μm), and the MIPS 24 μm band. The typical on-source integration time of the Spitzer observations was 10–30 minutes per source per band. The integration time for J0005−0006 at 15.6 and 24 μm was about four hours and for J0303−0019 at 24 μm was 50 minutes. The open circle in J0303−0019 indicates problematic photometry due to contamination from heavily blended neighbours. Circles with downward arrows represent 2 σ upper limits. Errors indicate measurement uncertainties (1 σ). Note that the IRAC 4.5 μm bandpass includes the strong Hα emission line (6,563 Å) at this redshift, enhancing the flux. The grey lines represent a template formed from the average of a large number of low-redshift type 1 quasar SEDs[13], and have been normalized to each object at λ = 3.6 μm. The blue and green lines represent the disk and hot dust components from the best model fits (the IRAC data are used to fit the disk emission, and the IRS PUI and MIPS data are then used to constrain the hot dust emission). The red lines show the sum of the two. In J0005−0006 and J0303−0019, a power-law is not a good description for the disk emission at $\lambda_0 > 1$ μm.

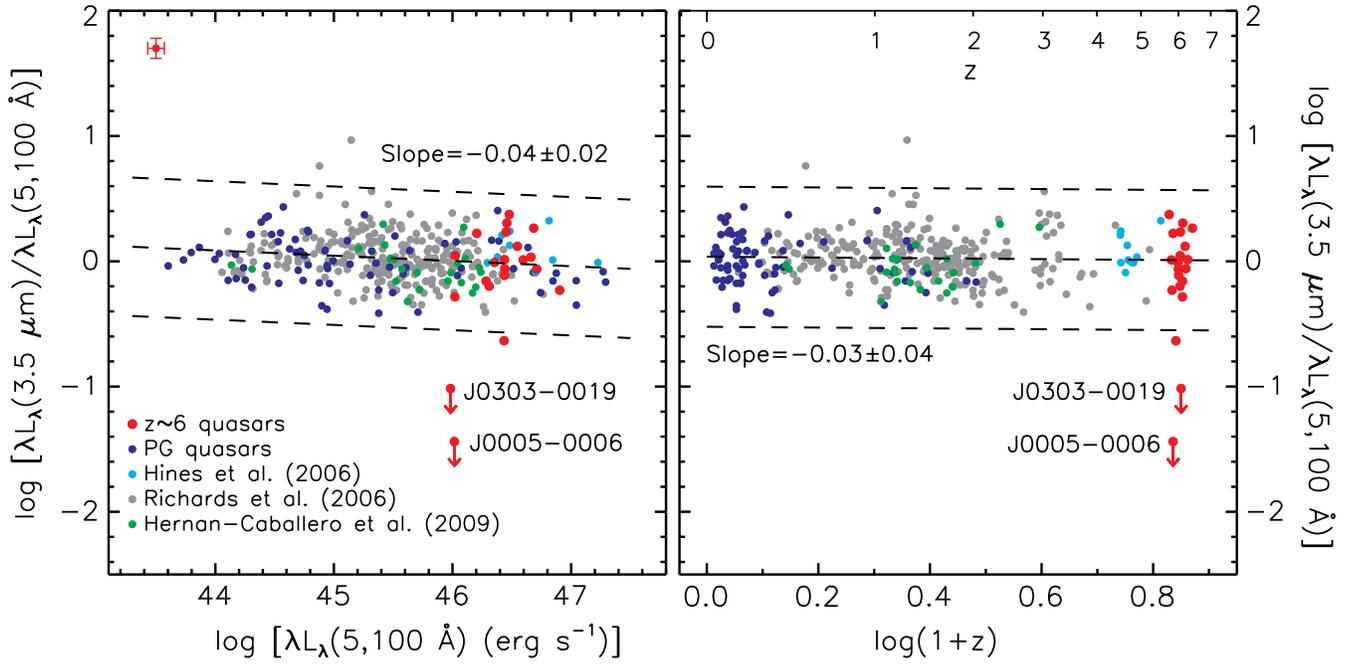

Figure 2. Luminosity and redshift dependence of the hot dust abundance for type 1 quasars. The left panel shows the hot dust abundance (i.e., rest-frame NIR-to-optical flux ratio) as a function of optical luminosity at rest-frame 5,100 Å. The right panel shows the hot dust abundance as a function of redshift. The red circles represent the 21 z ~ 6 quasars in our sample. Their rest-frame 5,100 Å and 3.5 μm luminosities are derived from the IRAC 3.5 μm and MIPS 24 μm fluxes, respectively. The other circles represent 362 low-redshift (0 < z < 5.5) quasars from the literature[12,13,19-22]. Typical errors (1 σ) for z ~ 6 quasars are given in the upper left corner of the left panel. In each panel, the dashed lines show the best linear fit to the sample and its 3 σ range.

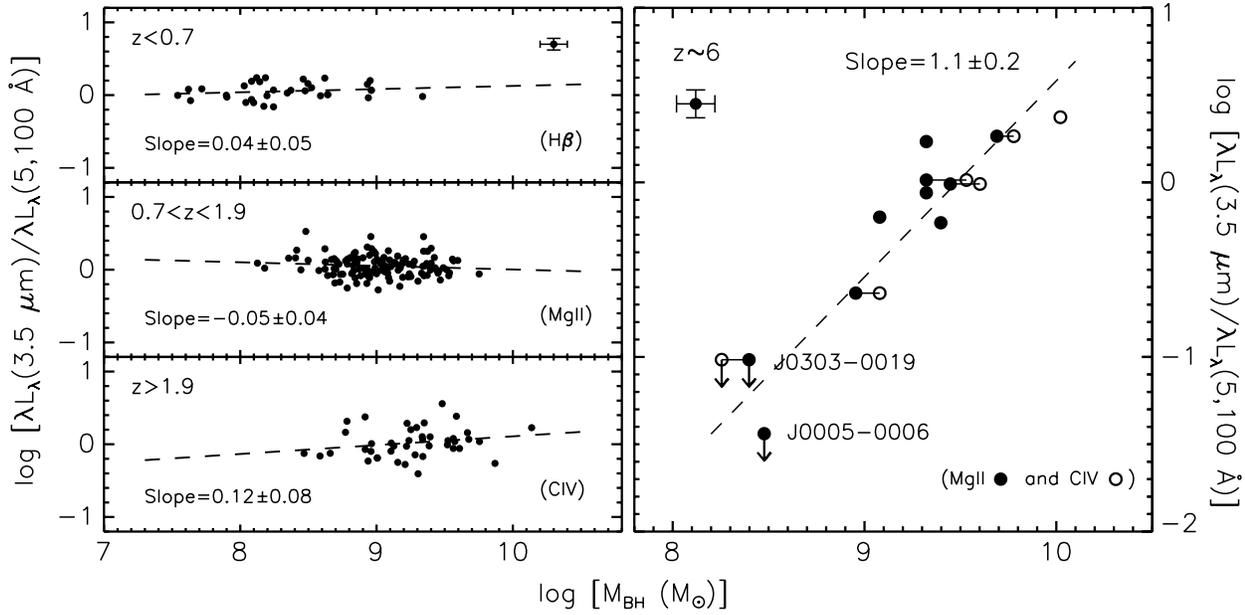

Figure 3. Correlation between the hot dust abundance and black hole mass for type 1 quasars. The left panels show a sample of 223 low-redshift SDSS quasars[13] whose black hole masses ($M_{BH}$) were derived from one of the three emission lines H$\beta$, MgII, and CIV[24] (ref. 27). Typical errors (1 $\sigma$) are given in the top panel. The dashed lines are the best linear fits. The right panel shows 11 z ~ 6 quasars with black hole mass measurements from MgII (filled circles) and/or CIV (open circles) in the literature[2,7,15,28]. Typical errors (1 $\sigma$) are given in the upper left corner of the panel. The circle pairs connected by short lines represent the same quasars measured with the two emission lines. All the mass values have been re-calibrated using the latest mass scaling relations[23,29]. The probability based on the generalized Kendall's $\tau$ test that a correlation between the hot dust abundance and black hole mass at z ~ 6 is present is 0.9987. The dashed line is the best linear fit using the Buckley-James regression method and the EM regression algorithm (they agree within the errors). The statistics were calculated within the Astronomy Survival Analysis software package[30]. If a quasar has two measurements from MgII and CIV, we take the average.